\documentclass[onecolumn,showkeys,preprintnumbers,aps,a4paper,amssymb,prd,superscriptaddress,nofootinbib]{revtex4-2}
\usepackage{comment}
\usepackage{graphicx}
\usepackage{epsf}
\usepackage{bm}
\usepackage{amsmath}
\usepackage{amsfonts}
\usepackage{amssymb}
\usepackage{epstopdf}
\usepackage{color}
\usepackage[dvipsnames]{xcolor}
\usepackage{verbatim}
\usepackage{multirow}
\usepackage{soul}
\usepackage{physics}
\usepackage{bm}

\usepackage[width=0.00cm, height=0.00cm, left=1.50cm, right=1.50cm, top=2.00cm, bottom=2.00cm]{geometry}
\usepackage{microtype}
\usepackage[colorlinks = true,
            linkcolor = red,
            urlcolor  = black,
            citecolor = red,
            anchorcolor = blue]{hyperref}
\usepackage[capitalize]{cleveref}
\usepackage[normalem]{ulem}
\usepackage{enumitem}
\usepackage{booktabs}

\usepackage{lipsum}

\makeatletter\let\expandableinput\@@input\makeatother

\begin{document}

\begin{center}
		\vspace{0.4cm} {\large{\bf Dynamical Oscillations in Dark Energy: Joint Constraints on the $w_{sin}$CDM Model  from DESI, OHD, and Supernova Samples}} \\
		\vspace{0.4cm}
		\normalsize{ Manish Yadav $^1$, Archana Dixit$^2$, M. S. Barak$^3$,  
         Anirudh Pradhan$^4$ }\\
		\vspace{5mm}
		\normalsize{$^{1,3}$ Department of Mathematics, Indira Gandhi University, Meerpur, Haryana 122502, India}\\
		\normalsize{$^{2}$ Department of Mathematics, Gurugram University, Gurugram- 122003, Harayana, India.}\\ 
        \normalsize{$^{4}$ Centre for Cosmology, Astrophysics and Space Science (CCASS), GLA University, Mathura-281 406, Uttar Pradesh, India}\\
		\vspace{2mm}
		$^1$Email address: manish.math.rs@igu.ac.in\\
		$^2$Email address: archana.ibs.maths@gmail.com\\
            $^3$Email address: ms$_{-}$barak@igu.ac.in\\
            $^4$Email address: pradhan.anirudh@gmail.com\\
\end{center}

{\noindent {\bf Abstract.}

In this study, we investigate the oscillatory dark energy model $w_{\sin}\mathrm{CDM}$ based on the DESI BAO data together with OHD, Pantheon Plus, and SH0ES measurements. We examine how the DESI data influence the dark energy 
equation-of-state plane $(w_0, w_a)$ within cosmological models that are free from  Hubble tension and employ a Monte Carlo Markov Chain (MCMC) approach. Our findings indicate that although the parameter space still favors $w_a < 0$ and $w_0 > -1$ , the cosmological constant remains consistent with the DESI+OHD+PP combination at the $2\sigma$ level. We also observe that the best-fit Hubble constant $H_0$ is higher for the DESI+OHD+PP+SH0ES data combination, leading to a residual Hubble tension of less than $1\sigma$  to remain consistent with the SH0ES measurement. These results suggest that attempts to address the Hubble tension tend to reduce indication of DESI for the oscillatory dark energy model. Therefore, claims that the cosmological constant should be approached with greater caution, considering both the latest observational datasets and the existing cosmological tensions. We also obtained the present deceleration parameter and the effective equation-of-state value as  $q_0 = -0.36$ and $w_{\mathrm{eff}} = -0.57$, respectively, for the DESI+OHD+PP+SH0ES dataset combination. Further analysis indicated a strong departure of $w_0$ from $w=-1$ at the $4\sigma$ level for the DR2+OHD+DES-5yr data combination. The inferred $\Omega_{m}$ tended to shift toward higher values when supernova samples were included, indicating a systematic preference for larger  $\Omega_{m}$ in combinations involving supernova data.

\section{Introduction}

The discovery of the universe accelerated expansion in 1998, based on observations of Type Ia supernovae (SNIa) \cite{ref1,ref2,ref3}, marked a major milestone in modern cosmology. Subsequent observational evidence has robustly confirmed this phenomenon, including data from Planck missions on the cosmic microwave background (CMB) and the Wilkinson Microwave Anisotropy Probe (WMAP) \cite{ref4,ref5,ref6}, large-scale structure (LSS) surveys such as  2dFGRS, SDSS, WMAP, and 6dFGS; baryon acoustic oscillation (BAO) measurements \cite{ref7,ref8,ref9,ref10,ref11,ref12}; as well as studies of weak gravitational lensing \cite{ref13,ref14,ref15}, galaxy clusters \cite{ref16} and high-redshift galaxies \cite{ref17}. In the same context of interpreting the accelerated expansion of the universe, several cosmological studies have attempted to modify the standard theory of gravity, as general relativity (GR) appears to face some challenges on cosmological scales \cite{ref18,ref19}. For a comprehensive discussion, refer to \cite{ref20}. Alternatively, other cosmological  researchers have proposed the existence of an unknown component with negative pressure,termed dark energy (DE), as the driving mechanism behind cosmic acceleration.\\

Over the past few decades, the $\Lambda$CDM (where $\Lambda$ and CDM are the cosmological constant and Cold Dark Matter, respectively) model has served as the standard framework in cosmology, which has an equation of state (EoS $w=-1$) and is remarkably consistent with observational cosmological astrophysical data across multiple scales and cosmic epochs \cite{ref1,ref2,ref21,ref22,ref23}.  However, it suffers significant fundamental discrepancies such as the Hubble tension (these tensions arise between indirect $H_0$ estimates from cosmological fits and direct local measurements using standard candles), the fine-tuning problem \cite{ref24} (observations indicate an extremely small cosmological constant $\Lambda$ driving cosmic acceleration, whereas quantum field theory predicts a value larger by nearly $10^{120}$), and the coincidence problem \cite{ref25} (which  points to similar magnitudes of $\Lambda$ and baryonic matter density).These discrepancies have motivated extensive efforts to explore alternative explanations, either by introducing new physical theories or by identifying possible systematic uncertainties in observations \cite{ref26, ref27, ref28, ref29}. Numerous studies have investigated novel physics as a potential solution to these cosmological tensions. Most adopt a bottom-up approach, assuming an effective field theory framework capable of explaining the constraints imposed by diverse and precise observational datasets.\\

Two primary methodologies have been suggested to explain the $H_0$ phenomenon:. The first keeps the foundation of general relativity intact but incorporates novel elements that extend beyond the current standard model of Particle Physics \cite{ref30}. The second methodology explores alternative gravity theories, such as  $f(T)$, $f(R)$ and $f(R,T)$ etc, introducing extra degrees of freedom and allowing for a different derivation of the Universe expansion rate. Under the common assumption that the universe behaves as a fluid, both approaches yield an expansion rate influenced by the evolution of the dark energy equation of state, denoted by $w$. The parameter $w$ is quantified as the ratio between the pressure ($p$) of the dark energy and its energy density ($\rho$), that is, $w=p/\rho$. The standard cosmological model of the universe posits that dark energy behaves as a fluid with constant negative pressure, characterized by an EoS parameter $w$ equal to $-1$ at all epochs,  meaning that the pressure-to-density ratio of dark energy remains fixed throughout the cosmic history. Conventionally, cosmologists often assume $w$ is a fixed value, not always exactly $-1$,\cite{ref31}  to simplify the  models. However, a more generalized approach expands the standard theory by permitting $w$ to evolve over cosmic time, rather than remaining constant. This possibility is typically addressed by defining $w$ as a function of the redshift, leading to different potential mathematical forms for its variation. Adopting these dynamic parameterizations gives cosmologists greater adaptability to fitting expansion histories to observational data, providing an improved match to how the universe behaves in reality.\\

Several studies have addressed the $H_0$ tension by assuming different forms of dark energy parameters, showing that the equation of state (EoS) deviates from $-1$. In Ref.\cite{ref32}, the authors considered the $w$CDM model (a one-parameter extension of dark energy) and obtained an $H_0$ measurement of $71.56\pm0.79$, corresponding to a $1.5\sigma$ deviation from the vacuum energy, using the combined dataset DESI BAO + BBN + OHD + SN + SH0ES. Other references \cite{ref33,ref34, ref35, ref36, ref37,ref38} have also investigated the $w$CDM framework. In Ref.\cite{ref39}, the authors assumed the CPL model (a two-parameter form of dark energy) and found $H_0 = 72.83 \pm 0.68$ with Planck18+DESI+Pantheon+SH0ES. Similar two-parameter parameterization approaches \cite{ref40,ref41,ref42,ref43,ref44}, such as the Logarithmic parametrization \cite{ref45,ref46,ref47} and the  Jassal-Bagla-Padmanabhan parameterization (JBP) model \cite{ref48,ref49}, have also been studied. Furthermore, high-dimensional extensions involving three and four free parameter forms of dark-energy models have been presented in studies \cite{ref50,ref51,ref52}. These parameterization forms have received significant attention, as recent studies suggest that observational data favor a dynamic dark energy equation of state. With the continuous improvement of precision cosmological observations, exploring various dark energy EoS parameterizations has become increasingly important.\\

 Recent studies have explored oscillating dark energy equations of state using a wide range of low- and high-redshift observational datasets, including CMB, BAO, CC, and  Pantheon supernova data, within several oscillatory dark energy parameterizations. In Ref. \cite{ref52a}, the authors investigated eight distinct oscillating dark energy models based on trigonometric (sine and cosine) functions. Their results indicate that none of these models show strong observational evidence for dynamic dark-energy behavior that departs significantly from the cosmological constant scenario. Furthermore, across all the considered oscillating models, the Hubble tension is not effectively resolved, with the inferred Hubble constant values remaining broadly consistent among the different parameterizations. In Ref. \cite{ref52b}, the authors examined $12$ different parametrizations of oscillating dark energy models using the BAO, CC, and PP data sets. These parameterizations included models with four free parameters and reduced two-parameter forms. Their analysis shows that the inferred values of the Hubble constant remain consistent within the range $H_0 \simeq 67\text{-}68\,\mathrm{kms^{-1}Mpc^{-1}}$. Moreover, they found that the dark energy EoS exhibits deviations from $\omega = -1$, indicating mild dynamical behavior, although these deviations are favored by observational data. The oscillatory dark energy models show no strong deviation of dark energy from the cosmological constant; therefore, these models are at most mildly favored over $\Lambda$CDM based on current observational data.  Motivated by this result, we adopt a specific functional form of oscillating dark energy (presented in Eq.\ref{eq6}) in order to further investigate whether such models can produce distinguishable signatures when confronted with recently released DESI BAO and superanova data (Union3, DES-5yr)observational constraints.\\

In this study, we introduced the oscillatory behavior of the dynamical dark energy model ($w_{\sin}$CDM), aiming to investigate its implications for the evolution of the universe. We examined the effect of parameters $w_0$ and $w_a$ on the evolution of other cosmological parameters from multiple combinations of observational datasets. We also studied important cosmological parameters such as the decleration parameter, total EoS parameter from the other and SH0ES combination. The remainder of this paper is organized as follows. In Section I, we outline  key developments and literature pertaining to the dynamic dark energy framework. Section II describes the observational datasets used in this analysis, namely, OHD, Pantheon Plus (PP), SH0ES, Union3, DES-5yr and DESI BAO—along with the adopted methodology.  Section III presents our results and discusses the implications of these findings. Section IV presents the key conclusions of this study.

\section{MODEL}\label{sec2}

We start from a spherically symmetric FLRW spacetime, and the line element takes the form:
\begin{equation}\label{eq1}
ds^2 = -c^2 dt^2 + a(t)^2 \left[ \frac{dr^2}{1 - \kappa r^2} + r^2 \left( d\theta^2 + \sin^2\theta \, d\phi^2 \right) \right],
\end{equation}

where $a(t)$ is the cosmic scale factor as a function of cosmic time $t$ and $\kappa$ denotes the cosmic curvature of the universe, which can take three numerical values: $0$ (flat universe), or $<0$(open universe), or $>0$ (closed universe). In the present study, we assume a flat universe, which is strongly supported by recent observational astrophysical data. By applying the FLRW metric, along with Einstein’s field equations and the energy–momentum tensor, we obtain the first and second Friedmann equations:
\begin{equation}\label{eq2}
	3 H^2= 8\pi G\rho
\end{equation}

\begin{equation}\label{eq3}
 2\dot{H} + 3H^2  =-\frac{8\pi G}{c^2}p
\end{equation}

In these equations, $\rho$ and $p$ denote the total energy density and total pressure of the entire cosmic fluid, respectively, and $H$ is the Hubble parameter, defined as $H = \dfrac{\dot a}{a}$ with $\dot a$ represents the time derivative of the scale factor. Each type of energy component (radiation/matter/dark energy) in the universe was independently conserved. Consequently, every component satisfies its own continuity equation, which represents the conservation of energy in an expanding universe and is derived from the first law of thermodynamics as follows: 

\begin{equation} \label{eq4}
 \dot{\rho_i}+3H(1+w_i)\rho_i=0 ,
\end{equation}

where subscript $i$ denotes each energy component of the Universe, such as radiation, matter, and dark energy, with the corresponding equation of state (EoS) parameter defined as $w_i = \dfrac {p_i}{\rho_i}$. For radiation, $w_r = 1/3$ then energy density of radiation becomes $\rho_r = \rho_{r0} a^{-4}$ and for matter, $w_m= 0$, giving the energy density of matter $\rho_m = \rho_{m0} a^{-3}$, where the subscript $0$ denotes the present value in the current universe. The dark energy density as obtained from the continuity equation, is expressed by Eq.\ref{eq5}:

\begin{equation}  \label{eq5}
\rho_{\mathrm{de}} = \rho_{\mathrm{de0}} \exp\left[-3 \int_{1}^{a} \frac{1 + w_{\mathrm{de}}}{a'} \, da'\right],
\end{equation}

In this study, we considered the oscillatory behavior of the dark energy equation of state (EoS) proposed by Zhang and Ma \cite{ref53}, which is expressed as

\begin{equation}  \label{eq6}
w_{de} = w_0 +w_a[a\sin(\dfrac{1}{a})-\sin1]
\end{equation}

From Eq.\ref{eq5} and Eq.\ref{eq6}, we get

\begin{equation}  \label{eq7}
\rho_{\mathrm{de}} = \rho_{\mathrm{de0}} \exp\left[-3 \int_{1}^{a} \frac{1 + w_0 +w_a[a'\sin(\dfrac{1}{a'})-\sin1]}{a'} \, da'\right],
\end{equation}

and the total density parameter satisfies the condition:

\begin{equation} \label{eq8}
\rho_{\mathrm{}} = \rho_r+\rho_m+\rho_{\rm{de}} 
\end{equation}

Substituting the energy densities of radiation, matter, and dark energy into Eq.\ref{eq8}, we obtained the oscillatory model ($w_{\rm{\sin}}$CDM) from the Friedmann equation Eq.\ref{eq2} in terms of the redshift $z$:

\begin{equation}
H(z)^2 =\dfrac{8\pi G}{3}\left[\rho_{r0} (1+z)^4 + \rho_{m0} (1+z)^3 + \rho_{\rm de0}(1+z)^{3(1+w_0-w_a\sin1)} \,
\exp\!\left( 3 w_a \int_0^z (\dfrac{\sin(1+z)}{(1+z)^2}) \, dz \right)
\right],
\end{equation}

where $\rho_{r0}$, $\rho_{m0}$ and $\rho_{de0}$ are the present values of radiation, matter and  dark energy  densities in the current universe, respectively.

\section{DATA AND METHODOLOGY}\label{sec3}


\textbf{Dark Energy Spectroscopic Instrument}:The first-year data release of the DESI Collaboration \cite{ref54,ref55} includes
seven tracers spanning different redshift intervals: "BGS ($0<z<0.4$), LRG1 ($0.4<z<0.6$), 
LRG2 ($0.6<z<0.8$), LRG+ELG ($0.8<z<1.1$), ELG ($1.1<z<1.6$), QSO ($0.8<z<2.1$), 
and Ly$\alpha$ QSO ($1.77<z<4.16$)". A summary of these redshift ranges, along with their associated comoving distances, is provided in Table~1 in Ref.~\cite{ref56}. Subsequently, the DESI Collaboration released the second data release (DESI DR2), which extended the analysis to nine tracers, as reported in Ref.\cite{ref56a}. These tracers allow for precise measurements of the BAO distance quantities 
$D_H(z)/r_{\rm d}$, $D_M(z)/r_{\rm d}$, and $D_V(z)/r_{\rm d}$, 
each evaluated at their corresponding effective redshift, defined as:

\begin{itemize}
\item $\textbf{Transverse Comoving Distance:}$\\

$D_{M}(z)=  \int_0^z  {c \over H(z')} \text{d}z'$, where $c$ is the speed of light. \\

\item\textbf{Hubble Distance}:
$ D_H(z) = \frac{c}{H(z)}$,

\item\textbf{Angle-Averaged  Distance}:
$D_V(z) \equiv \left[z D^2_M(z) D_H(z)\right]^{1/3}$ ,
 
\end{itemize}
where $r_{\rm d}=\int_{z_{\rm d}}^\infty \frac{c_{\rm s}\text{d}z}{H(z)}$  and  ($c_{\rm s}$) are the sound horizon at the drag redshift ($z_{\rm d}$) and the sound speed  of the baryon-fluid of the universe, respectively .\\

\textbf{Observational Hubble Data}:The Observational Hubble Data (OHD) technique determines the expansion rate of the Universe by measuring the differential ages of the oldest passively evolving galaxies located near redshifts. This method offers a model–independent estimate of the Hubble parameter by connecting redshift and cosmic time through the relation 
\[
H(z) = -\frac{1}{1+z}\,\frac{dz}{dt},
\]
as first introduced in early foundational works~\cite{ref57}. 
In this analysis, we employ 33 independent measurements of $H(z)$ spanning the range 
$0.07 \leq z \leq 1.965$, as reported in 
Refs.~\cite{ref58,ref59,ref60,ref61,ref62,ref63,ref64,ref65}, 
incorporating data from a variety of Hubble parameter surveys.

We define the chi-squared function for the observational Hubble data, denoted by $\chi^2_{\rm OHD}$, as follows:

$$\chi^2_{\text{OHD}} = \sum_{i=1}^{33} \frac{\left[ d^{obs}(z_i) - d^{th}(z_i) \right]^2}{\sigma^2_{d^{obs}(z_i)}},$$

Where $ d^{obs}(z_i)$ and $ d^{th}(z_i)$ denotes the observed and model-predicted values of the Hubble parameter at redshift $z_i$, respectively, and $\sigma_{{d^{obs}(z_i)}}$ represents the associated observational uncertainty. In Fig.\ref{fig1}, our model and the $\Lambda$CDM model nearly overlap across the entire redshift range, indicating that our model is strongly consistent with both OHD data and the $\Lambda$CDM model. \\

\begin{figure}[hbt!]
    \centering
    \includegraphics[width=0.8\linewidth]{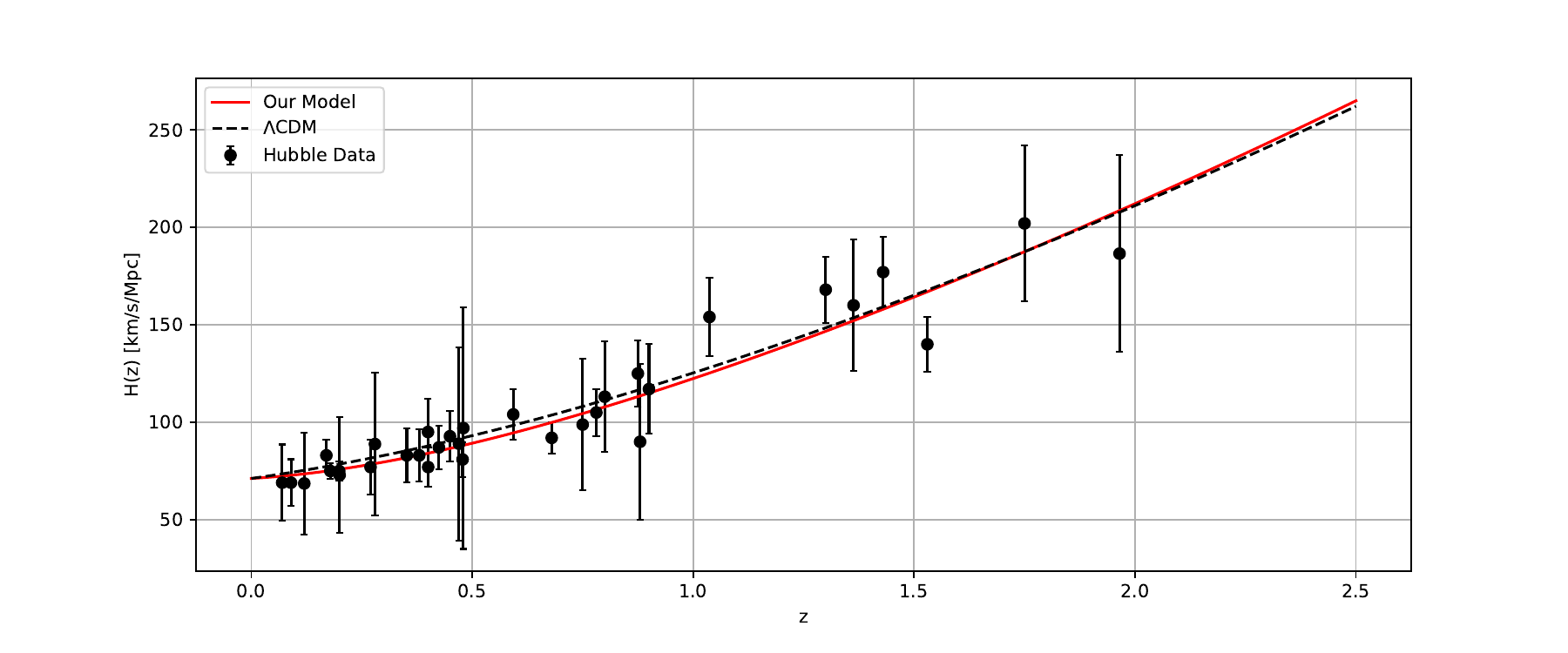}
    \caption{The 2D plot between $H(z)$ verse $z$ for our model and $\Lambda$CDM model from $33$ OHD data points with black error bar. }
    \label{fig1}
\end{figure}
\begin{figure}[hbt!]
    \centering
    \includegraphics[width=0.8\linewidth]{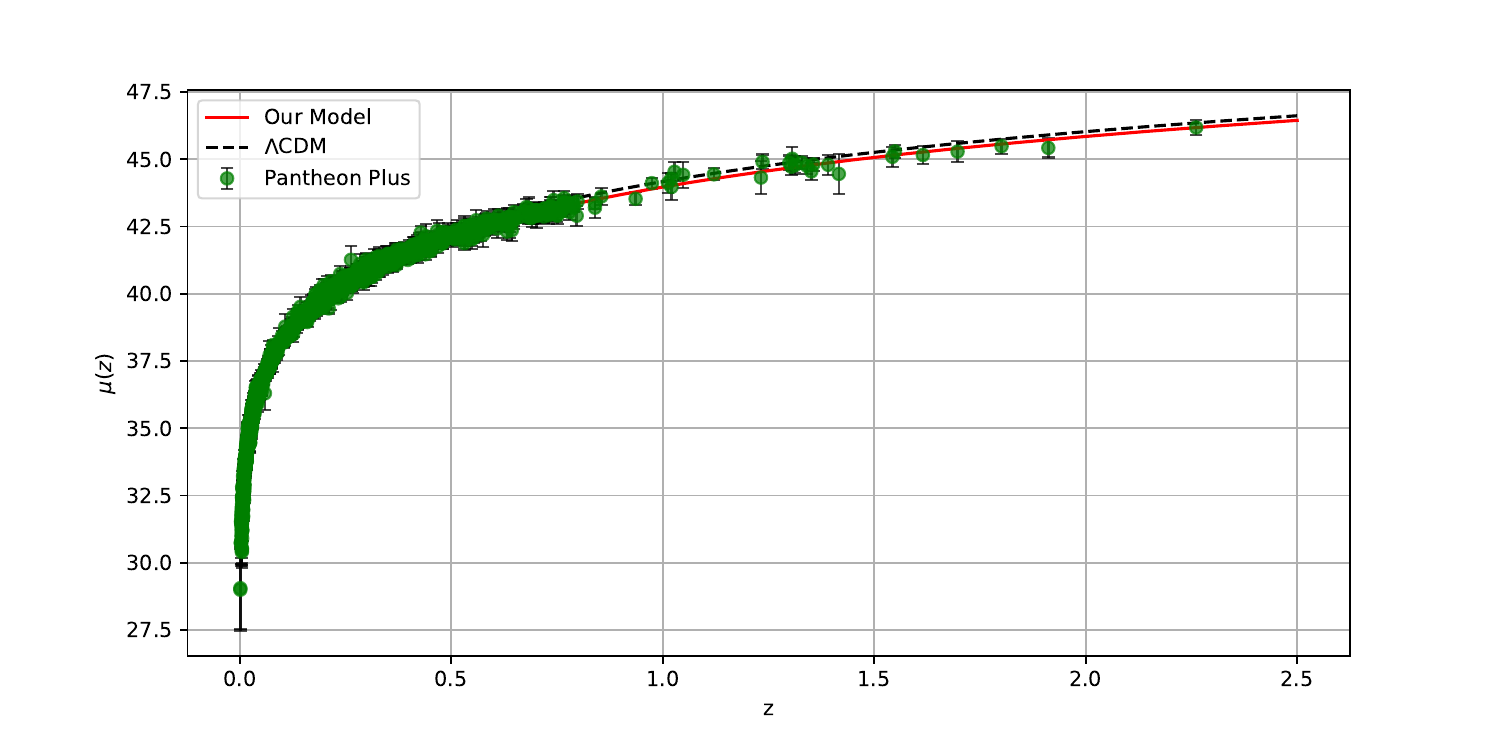}
    \caption{The 2D plot between $\mu(z)$ verse $z$ for our model and $\Lambda$CDM model from $1701$ PP data points with green error bar.}
    \label{fig2}
\end{figure}
\textbf{Pantheon Plus\&SH0ES}: 
The primary evidence for accelerated cosmic expansion came from Type Ia supernovae (SNIa) measurements.  SNIa are among the most dependable and effective probes for examining the nature of dark energy and the cosmic expansion. Several excellent supernova datasets have been assembled over time; the most recent development is the revised Pantheon Plus sample.  The 1701 light curves in this compilation, which span a broad redshift range of $ z \in [0.001, 2.26]$ \cite{ref66}, correspond to 1550 different SNe Ia events.  We also include the newest SH0ES Cepheid-calibrated distances \cite{ref67} in our likelihood analysis.  Our model agrees well with the PP data and closely follows the $\Lambda$CDM curve, shown in Fig.\ref{fig2}.\\

\textbf{Union3}: We use the Union3 compilation, a comprehensive dataset consisting of 2087 Type~Ia supernovae drawn from 24 independent samples, covering the redshift range $0.01 < z < 2.26$. All supernovae were standardized to a consistent distance scale using the SALT3 light-curve fitter. In this analysis, we employed the 22 binned distance moduli of the Union3 sample, as reported in Ref.~\cite{ref67a}.\\

\textbf{DES-5yr}: We used a DES-5yr sample \cite{ref67b} consisting of 1829 Type~Ia supernovae spanning the redshift range $0.025 < z < 1.13$. This dataset was drawn from the five-year Dark Energy Survey supernova program and provides one of the largest homogeneous samples of spectroscopically and photometrically classified SNe~Ia to date. The supernovae were uniformly calibrated and standardized using state-of-the-art light-curve fitting techniques, making the DES-5yr sample a powerful probe of the cosmic expansion history.\\

In this study, we employ the Boltzmann  CLASS  code \cite{ref68} together with the Bayesian analysis framework \texttt{MontePython} \cite{ref69} to derive cosmological constraints using the Markov Chain Monte Carlo (MCMC) technique. We aimed to explore the oscillatory behavior of the dynamical dark energy model ($w_{\sin}$CDM). The  MCMC chains result are analyzed within GetDist \cite{ref70}, and the convergence of the chains is ensured by applying the Gelman–Rubin criterion \cite{ref71}, requiring $R - 1 \leq 0.01$. For each parameter constraint, we run four independent chains with 100,000 steps, discarding the first 30,000 steps as burn-it. We adopted uniform priors for the cosmological parameters as follows: the baryon density $\Omega_b h^2 \in [0.005, 0.1]$, the cold dark matter density $\Omega_c h^2 \in [0.001, 0.99]$, the Hubble constant $H_0 \in [20, 100]$, the present dark energy equation of state $w_0 \in [-3, 1]$, and its scale-dependent evolution parameter $w_a \in [-3, 2]$.

\section{Results and discussion}

In this study, we analyze both the free ($\omega_b, \omega_{c}, H_0, w_o, w_a$) and derived ($\Omega_m, t_0$) cosmological parameters within the $w_{\text{sin}}$CDM and  $\Lambda$CDM models, using various combinations of observational datasets, including DESI+OHD, DESI+OHD+PP, and DESI+OHD+PP+SH0ES. We begin by discussing the behavior of the dark-energy EoS parameters $w_0$ and $w_a$, as they play a central role in characterizing the dynamical properties of dark energy within the $w_{\text{sin}}$CDM scenario.

\begin{figure}[hbt]
	\centering
	\includegraphics[width=0.45\linewidth]{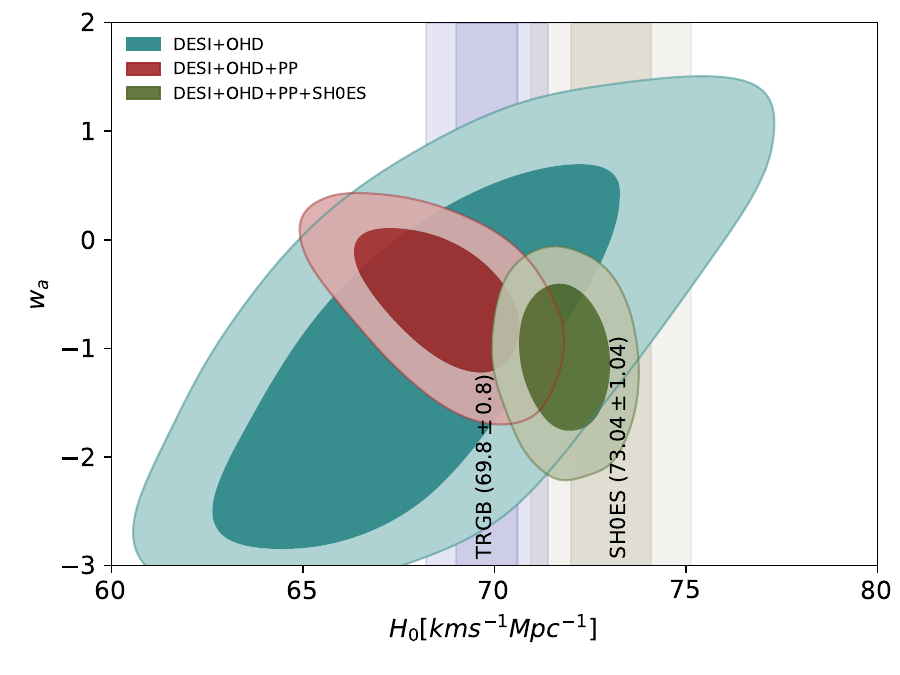}
	\includegraphics[width=0.45\linewidth]{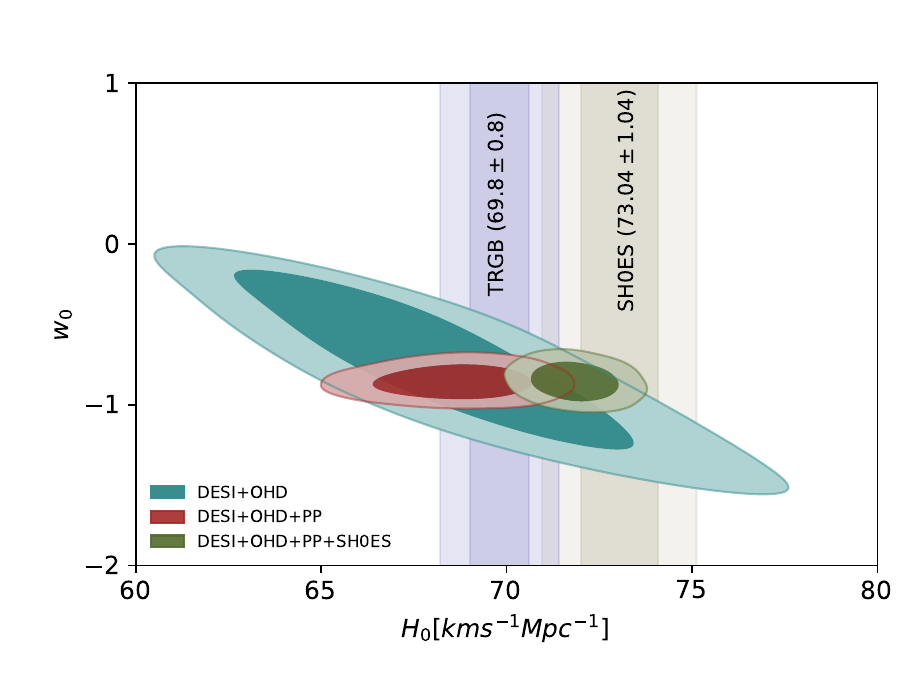}
	\caption{In this figure, the left panel shows the 2D contour between $H_0$ and $w_a$ overlaid with SHOES and TRGB bands, while  right panel displays 2D contour between $H_0$ and $w_0$ with the same external bands. These contour obtained from $w_{sin}$CDM model based on all the given dataset combinations is presented in  first row of Table \ref{tab1}.  }
	\label{fig4}
\end{figure}

\begin{table*}[hbt!]
	\caption{ The marginalized constraints, presented as mean values with 68$\%$ confidence levels (CL), on both the free and select derived parameters of the $w_{sin}$CDM and $\Lambda$CDM models for various datasets combinations.}
	\label{tab1}
	\scalebox{0.85}{
		\begin{tabular}{lccc}
			\hline
			\toprule
			\textbf{Dataset }&\;\;\;\;\;\textbf{DESI+OHD}\;\;\;\;\;& \textbf{DESI+OHD+PP} \;\;\;\;\;\;& \;\;\;\textbf{DESI+OHD+PP+SH0ES} 
			\\ \hline
			\textbf{Model} & \textbf{$\bm{w_{sin}}$CDM}\,&\textbf{$\bm{w_{sin}}$CDM}\,&\textbf{$\bm{w_{sin}}$CDM}\vspace{0.1cm}\\
			&\textcolor{teal}{\textbf{$\bm{\Lambda}$CDM}}\, & \textcolor{teal}{\textbf{$\bm{\Lambda}$CDM}}\, & \textcolor{teal}{\textbf{$\bm{\Lambda}$CDM}} 
			\\ \hline

			{\boldmath$10^{2}\omega_{b}$}&{$2.266\pm 0.037$ }&$2.269\pm 0.037$&$2.277\pm 0.037$\\
			
			&\textcolor{teal}{$2.271\pm 0.037$} &\textcolor{teal}{$2.267\pm 0.037$} &\textcolor{teal}{$2.298\pm 0.037$}\\

			{\boldmath$\omega{}_{\rm cdm }$}&$0.125^{+0.0110}_{-0.0095} $ &$0.122^{+0.0110}_{-0.0098} $ &$0.1429\pm 0.0072$\\
			
			&\textcolor{teal}{$0.1179^{+0.0062}_{-0.0069}$} &\textcolor{teal}{$0.1251\pm 0.0059 $} & \textcolor{teal}{$0.1327\pm 0.0059$}\\

			{\boldmath$H_0$ [$\text{km} \text{s}^{-1} \text{Mpc}^{-1}$]}&$68.30^{+2.90}_{-3.90}$&$ 68.60^{+1.50}_{-1.30} $& $71.85\pm 0.79$\\
			
			& \textcolor{teal}{$69.41\pm 0.67 $}&\textcolor{teal}{$69.36\pm 0.63 $}  & \textcolor{teal}{$70.97\pm 0.56$}\\

			{\boldmath$w_{0}$} &$-0.74^{+0.40}_{-0.31}$ &$-0.86\pm 0.07$ &$-0.85\pm 0.08 $ \\

			&-& \textcolor{teal}{-}& \textcolor{teal}{-}\vspace{0.2cm} \\
			{\boldmath$w_{a}$} &$-1.03^{+0.98}_{-1.40}$ &$-0.54^{+0.48}_{-0.39}$ &$-1.09^{+0.47}_{-0.42} $ \\
			
			&-& \textcolor{teal}{-}& \textcolor{teal}{-}\\

			{\boldmath$M_B{\rm[mag]}$}&-  &$-19.388^{+0.048}_{-0.039} $&$-19.292\pm 0.022$ \\
			
			&- &$-19.378\pm 0.022$& \textcolor{teal}{$-19.330\pm 0.017$}\\

			{\boldmath$\Omega{}_{m }  $}&$0.319^{+0.041}_{-0.035}   $ &$ 0.308\pm 0.013 $&$0.321\pm 0.011 $\\
			
			&\textcolor{teal}{$0.292\pm 0.013$}& \textcolor{teal}{$0.307\pm 0.011 $} & \textcolor{teal}{$0.309\pm 0.011 $}  \\

			{\boldmath$t{}_{0}  $}&$13.53^{+0.25}_{-0.30}$ &$ 13.59^{+0.26}_{-0.31}$  &$ 13.01\pm 0.17          $\\
			
			&\textcolor{teal}{$13.69\pm 0.21$}& \textcolor{teal}{$13.50\pm 0.19$} & \textcolor{teal}{$13.17\pm 0.17  $}  \\

			\hline

			{\boldmath$ \chi^{2}_{min}$}&$ -14.08$&$-718.88$&$-662.82$ \\
			
			& \textcolor{teal}{$-14.33$}& \textcolor{teal}{$-721.05$} &\textcolor{teal}{$-666.22$}\\

			{\boldmath$ \text{AIC}$}&$ 24.08$&$728.88$&$672.82$\\
			
			& \textcolor{teal}{$20.33$}& \textcolor{teal}{$727.05$} &\textcolor{teal}{$672.22$}\\

			{\boldmath$\Delta \text{AIC}$}&$3.75$&$1.83$&$0.60$\\
			
			

			\hline
			\hline
		\end{tabular}
	}
\end{table*}

In this section, we present an analysis of the dark energy EoS parameters within  $w_{\text{sin}}$CDM based on joint constraints from multiple observational datasets. As shown in Table \ref{tab1}, the current-day EoS of dark energy obtain $w_{0} = -0.74^{+0.40}_{-0.31}$ for $w_\text{sin}$CDM with DESI+CC data sets. It is also observed that the upper bound on the time-dependent component of the dark energy parameter $w_a$ is well-constrained, whereas the lower bound remains unconstrained. Consequently, the 1D posterior distribution of $w_a$ obtained from the DESI+CC dataset alone did not exhibit a well-defined peak as shown in Fig.\ref{fig8}. The present numerical value of $w_0$ indicates the presence of the quintessence foam of dark energy and its deviation from the cosmological constant is $0.7 \sigma$. For the DESI+OHD+PP dataset combination, we obtain $w_{0} = -0.86 \pm 0.07$, which again suggests a present-day quintessence behavior and shows a $2.0\sigma$ deviation from $w = -1$. In contrast, the time-evolving parameter was constrained as $w_{a} = -0.54^{+0.48}_{-0.39}$, further supporting a mildly dynamical quintessence form of dark energy. When the SH0ES prior is included (DESI+OHD+PP+SH0ES), we find $w_{0} = -0.85 \pm 0.08$, indicating a similar quintessence-like trend with a $1.9\sigma$ deviation from the cosmological constant and time-varying dark energy component constrain $w_{a}= -1.09^{+0.47}_{-0.42}$. This behavior suggests a possible transition into the phantom regime ($w < -1$) in the past, providing further support for dynamical dark energy rather than a constant-$w$ model. As shown in Fig.\ref{fig8}, the parameters $w_0$ and $w_a$ display a pronounced negative correlation, with the contour regions demonstrating that the combined datasets tightly constrained both the parameters.\\

One of the most prominent and persistent challenges in contemporary cosmology is the so-called Hubble tension, which refers to the significant discrepancy in the inferred values of the Hubble constant, $H_0$, obtained from early- and late-Universe observations. Local distance-ladder measurements by the SH0ES collaboration yielded a high value of $H_0 = 73.04 \pm 1.04~\mathrm{km\,s^{-1}\,Mpc^{-1}}$ \cite{ref67}, based on Cepheid-calibrated Type~Ia supernovae. In contrast, the Planck collaboration’s analysis of the cosmic microwave background within the $\Lambda$CDM framework yields a substantially lower value, $H_0 = 67.36 \pm 0.54~\mathrm{km\,s^{-1}\,Mpc^{-1}}$ \cite{ref21}. This mismatch, now reaching a statistical significance of approximately $5\sigma$—has become a central puzzle in modern cosmology. In this section, we report the constraints on the key cosmological parameters, the Hubble constant, the matter density parameter, and the age of the universe for the $w$CDM and $\Lambda$CDM models using various observational datasets. The Hubble constant values obtained from the different dataset combinations show consistent cosmological evolution. Using the DESI+OHD dataset, we obtain $H_{0} = 68.30^{+2.90}_{-3.90}(69.41 \pm 0.67$) for the $w$CDM ( $\Lambda$CDM) models, both of which agree well with independent Planck (TRGB) measurements. For the same dataset combination, we find slightly higher values of $\Omega_{m} = 0.319^{+0.041}_{-0.035}$ for $w$CDM and also obtain $\Omega_{m} = 0.292 \pm 0.013$ for $\Lambda$CDM, while the inferred ages of the Universe remain consistent with the Planck-calibrated \cite{ref21} values for both consider models. After adding the PP dataset into the first given combination, we observe that the mean values of $H_{0}$ and $\Omega_{m}$ remain nearly unchanged for both models, but their $1\sigma$ uncertainties reduce, leading to tighter constraints on these parameters. Further analyze our model and standard model from DESI+OHD+PP+SH0ES, we obtain for our model a mean value of $H_0 = 71.85 \pm 0.79~\mathrm{km\,s^{-1}\,Mpc^{-1}}$, $\Omega_m = 0.321 \pm 0.011$, and an age of the Universe $t_0 = 13.01 \pm 0.17~\mathrm{Gyr}$. For the standard model, the corresponding values are $H_0 = 70.97 \pm 0.56~\mathrm{km\,s^{-1}\,Mpc^{-1}}$, $\Omega_m = 0.309 \pm 0.011$, and $t_0 = 13.17 \pm 0.17~\mathrm{Gyr}$. We compare these Hubble constant values obtained from both models with the SH0ES measurement. We find that our model retains the Hubble tension at approximately a $0.9\sigma$ level, whereas the $\Lambda$CDM model retains a Hubble tension of approximately $1.8\sigma$ level. Consequently, our model offers a slightly better agreement than the standard model. From the left and right panels of Fig.\ref{fig4}, we observe that for the DESI+OHD+PP+SH0ES dataset, the contour regions of our model overlapped well with the SH0ES band, indicating that the model is properly constrained and showed good consistency with the SH0ES measurement.\\

\textbf{Akaike Information Criterion}: In this work, we perform the Akaike Information Criterion (AIC)~\cite{ref72} to compare the $w_{\text{sin}}$CDM models based on their fit to different combinations of observational data such as DESI+CC, DESI+CC+PP, and DESI+CC+PP+SH0ES, where $\mathrm{AIC} = \chi^2_{\text{min}} + 2M$, and $M$ is the number of model-independent free parameters considered. To clearly illustrate the differences in AIC values between the $\Lambda$CDM and $w_{\text{sin}}$CDM models, we adopted the $\Lambda$CDM model as the reference with the smallest AIC value, then we set $\Delta \rm{AIC}_{{\Lambda}\rm{CDM}}$ to zero. The resulting values of $\Delta \mathrm{AIC}$ are presented in the last row of Table \ref{tab1}, where $\Delta \mathrm{AIC} = \mathrm{AIC}_{w_{\text{sin}}\mathrm{CDM}} - \mathrm{AIC}_{\Lambda \mathrm{CDM}}$.

\begin{figure}[hbt!]
    \centering
    \includegraphics[width=0.7\linewidth]{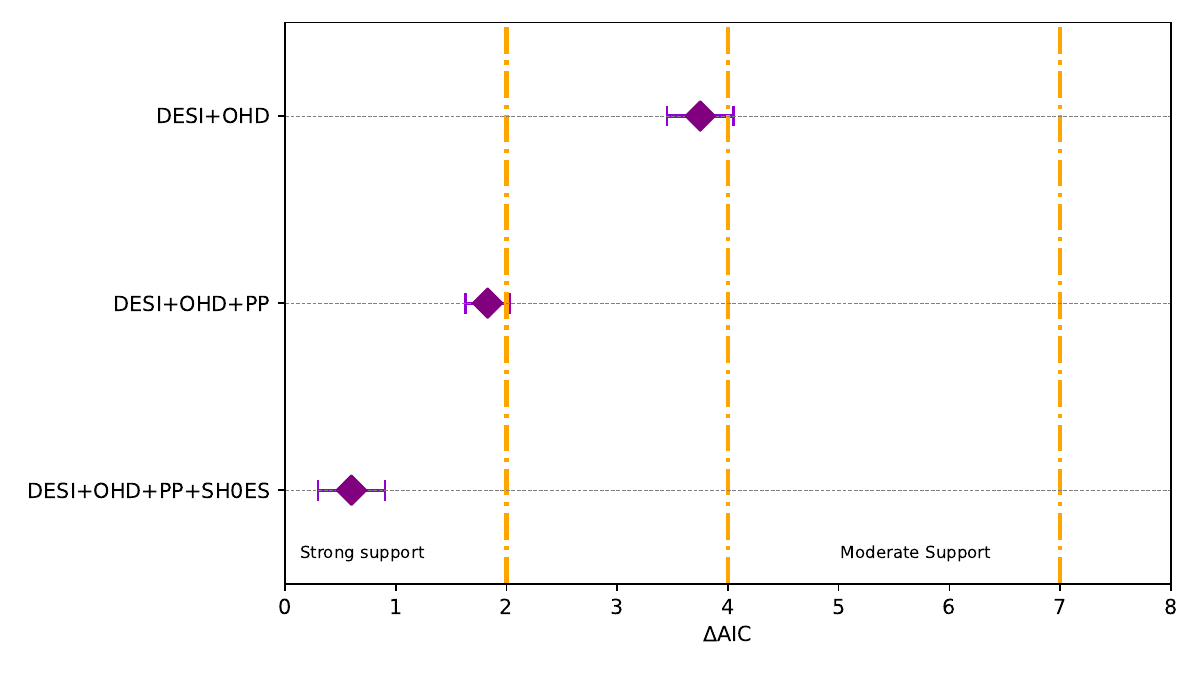}
    \caption{The $\Delta$AIC value of our model relative to the standard $\Lambda$CDM model for all the given dataset combinations is presented in  first row of Table \ref{tab1}. }
    \label{AIC}
\end{figure}

Fig.\ref{AIC} clearly shows that when using the DESI+OHD dataset, our model falls nearly within the region of moderate observational support, indicating reasonable consistency with the data. However, when additional datasets, such as PP and SH0ES, are included—specifically in the DESI+OHD+PP and DESI+OHD+PP+SH0ES combinations—the level of agreement between our model and the observational data becomes significantly stronger, as clearly shown in Fig.\ref{AIC}. This enhanced consistency demonstrates that our model performs effectively with a combination of astrophysical cosmological observations data and outperforms the standard $\Lambda$CDM model in explaining current data trends. \\

\textbf{Deceleration Parameter}: The deceleration parameter, denoted $q(z)$, is a key measure in cosmology that quantifies how the expansion of the universe accelerates or decelerates with redshift for our oscillatory dynamical dark energy model. In contrast, a negative value of the deceleration parameter, $q<0$, reflects an accelerating phase of cosmic expansion, first revealed through the late $20$th-century discovery of dark energy. However, a positive $q > 0$ value implies a decelerating universe (the expansion rate of the universe slows down), whereas $q = 0$ describes a critical equilibrium point where the expansion proceeds uniformly in cosmic time, with neither acceleration nor deceleration. This parameter plays a crucial role in defining the ultimate fate of the universe and provides valuable insights into its dynamics, particularly the influence of dark energy. \\

\begin{figure}[hbt!]
    \centering
    \includegraphics[width=0.6\linewidth]{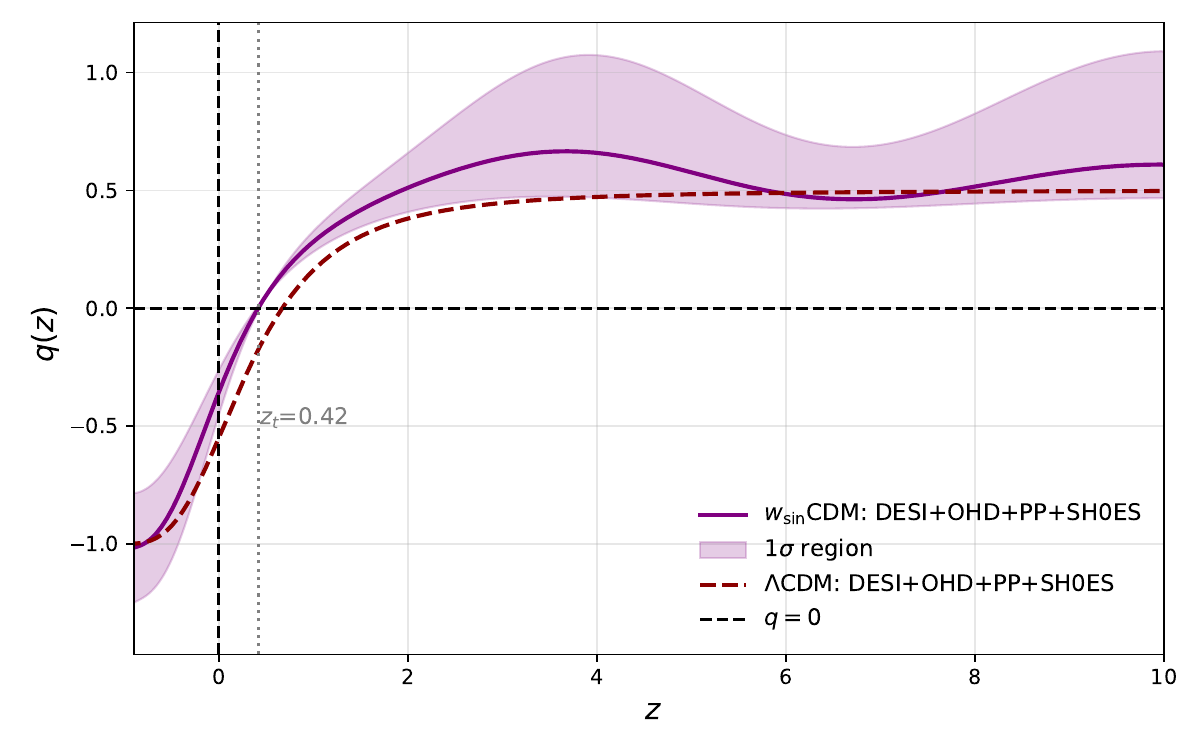}
    \caption{The 2D plot between $q(z)$ verse z for our model and $\Lambda$CDM model from DESI+OHD+PP+SH0ES data sets with 1$\sigma$ error bar.}
    \label{qz}
\end{figure}

As shown in Fig.\ref{qz}, the deceleration parameter $q(z)$ (depicted by the purple curve with $1\sigma$ error bars) is derived from our model using the combined DESI+OHD+PP+SH0ES datasets. Fig.\ref{qz} shows a horizontal black dotted line representing $q=0$ and a vertical black dotted line corresponding to $z=0$. We find the current deceleration parameter value to be $q_0 = -0.36 (-0.55)$   and the transition redshift to be $z_{\mathrm{tr}} \approx 0.42(0.66)$ from our model and standard model, respectively, with DESI+OHD+PP+SH0ES datasets, indicating the epoch at which the Universe transitions from a decelerating to an accelerating expansion phase. Furthermore, the analysis was extended to include  negative redshift values suggests that the cosmic expansion rate will grow exponentially in the future. Our obtained transition redshift and present-day $q_0$ for both model are in close agreement with recent results reported in the literature \cite{ref73,ref74,ref75,ref76}.\\

\textbf{Effective Equation of State}: The equation of state (EoS) parameter, $w_\text{eff}$, plays a crucial role in distinguishing between the different epochs of accelerated and decelerated expansion in the universe.\\

The evolution of the universe can be categorized into four major phases based on the effective equation of the state parameter. The stiff fluid phase ($w_{\text{eff}} = 1$) corresponds to a hypothetical state in which the pressure equals the energy density. This was followed by the radiation-dominated era ($w_{\text{eff}} = 1/3$), during which radiation governed the energy content of the universe immediately after the Big Bang. Subsequently, a matter-dominated phase ($w_{\text{eff}} = 0$) emerged, where non-relativistic matter became dominant, leading to the formation of galaxies and large-scale structures. Finally, the Universe enters the accelerated expansion phase ($w_{\text{eff}} < -1/3$), characterized by a negative equation of state parameter that causes repulsive gravity, resulting in the present acceleration driven by dark energy.\\

\begin{figure}[hbt!]
    \centering
    \includegraphics[width=0.6\linewidth]{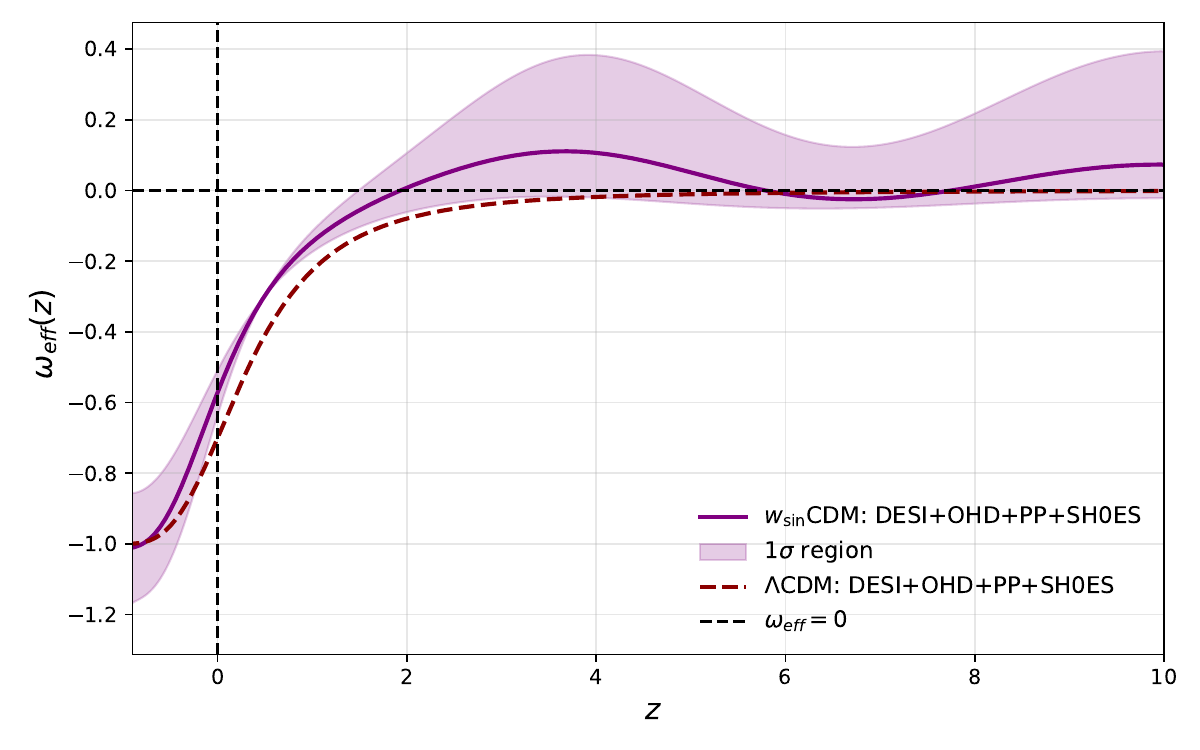}
    \caption{The 2D plot between $w_{\text{eff}}(z)$ verse z for our model and $\Lambda$CDM model from DESI+OHD+PP+SH0ES data sets with 1$\sigma$ error bar.}
    \label{weff}
\end{figure}

As shown in Fig.\ref{weff}, the EoS parameter $w_\text{eff}(z)$ (depicted by the purple curve with $1\sigma$ error bars) was derived for the $w_{sin}$CDM model based on the combined DESI+OHD+PP+SH0ES datasets. The figure includes a horizontal and vertical black dotted lines representing $q=0$ and  $z=0$, respectively. In contrast, we find the current EoS parameter value to be $w_\text{eff} = - 0.57$ ($-0.70$) from our model and the standard model, respectively, using the DESI+OHD+PP+SH0ES datasets,  which indicates the present-day quintessence nature of dark energy.\\

\begin{center}
	\textbf{Analysis $w_\text{sin}$CDM model with Supernova data(Union3, PP, DES-5yr)}
\end{center}

Furthermore, we perform a comprehensive analysis of the proposed model ($w_{\rm sin}$CDM) and CPL  ($w_0w_a$CDM) cosmological models by combining the DR2 and OHD data sets with several contemporary Type~Ia supernova compilations, namely Union3, PP, and DES-5yr. This multi--data-set approach allows us to robustly test the viability of the models and to assess the impact of different supernova samples on the inferred cosmological parameters. All free and derived parameters are tightly constrained and summarized in Table \ref{tab2}, while their corresponding two-dimensional marginalized confidence regions at the 68\% and 95\% C.L are presented in Fig.\ref{fig3a}, illustrating the correlations and degeneracies among the model parameters. From the DR2+CC+Union3 data combination, we obtain constraints on the Hubble constant of
$H_0 = 66.8 \pm 1.2\,\mathrm{km\,s^{-1}\,Mpc^{-1}}$ for the $w_{\rm sin}$CDM model, and
$H_0 = 66.7 \pm 1.1\,\mathrm{km\,s^{-1}\,Mpc^{-1}}$ for the standard $w_0w_a$CDM scenario. These results are mutually consistent and lie on the lower side of the local distance-ladder measurements, while remaining compatible with several late-time probes \cite{ref71a}.\\

\begin{figure}[hbt!]
	\centering
	\includegraphics[width=0.8\linewidth]{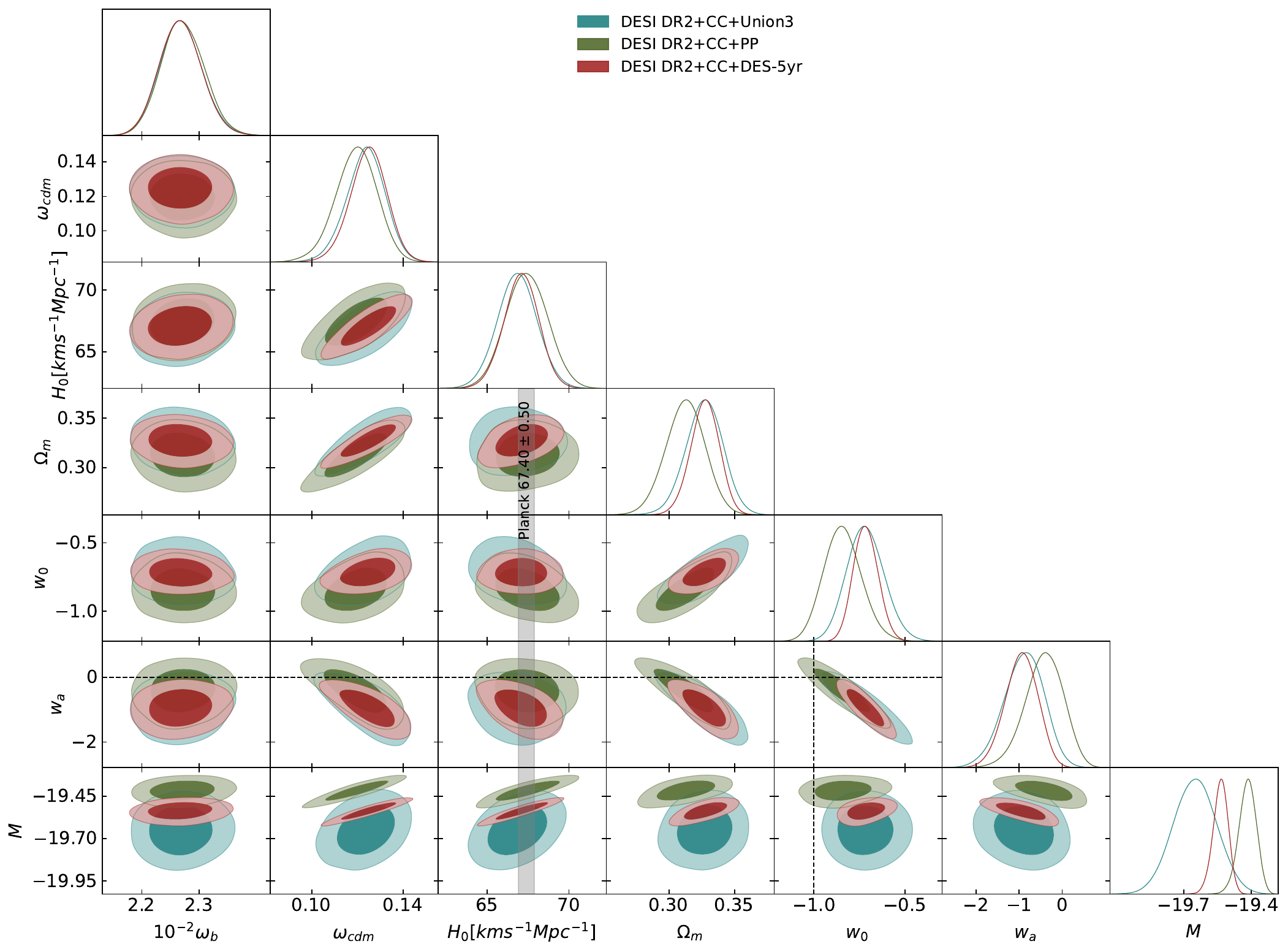}
	\caption{}
	\label{fig3a}
\end{figure}

We further find that the present-day value of the dark energy EoS is approximately $w_0 \simeq -0.73$ for both considered cosmological models, as reported in Table \ref{tab2}. This value deviates from the cosmological constant case ($w=-1$) at roughly the $2.8\sigma$ C.L, thereby indicating a statistically significant preference for a dynamical dark energy component within the framework of the adopted data sets. In addition, the dynamical parameter governing the redshift evolution of the EoS is constrained to be negative, with $w_a < 0$, corresponding to a deviation at the $\sim 2\sigma$ level. This behavior suggests a time-evolving dark energy sector that departs from a constant equation of state at late times. The matter density parameter is consistently constrained to relatively higher values, with $\Omega_m \simeq 0.325$ for both the $w_{\rm sin}$CDM and $w_0w_a$CDM models. This result reflects the interplay between the supernova, cosmic chronometer, and DR2 data sets, and highlights the role of combined late-time observations in tightening the constraints on the background cosmological parameters.\\

From the combined DR2+CC+PP data set, we find constraints on the Hubble constant of  $H_0 = 67.5 \pm 1.3$($67.7 \pm 1.2$)$\mathrm{km\,s^{-1}\,Mpc^{-1}}$ for the $w_{\rm sin}$CDM ($w_0w_a$CDM ) models. These mean value for our and CPL models are in good agreement with Planck calibration, meanwhile these value have slightly large error bar ($>1\sigma$). The present-day  EoS parameter $w_0$ shows a deviation from $w=-1$ at the $1.6\sigma$($w_{\rm sin}$CDM) and  $2.0\sigma$($w_0w_a$CDM) C.L, whereas the dynamic parameter $w_a$ deviates from zero at approximately the  $1\sigma$ C.L. These results consistently indicate the presence of dynamic dark energy in both models. Moreover, the $\Omega_m$ was found to have nearly the same mean value for both the models, as summarized in Table \ref{tab2}.\\ 

\begin{table*}[hbt!]
	\caption{ The marginalized constraints, presented as mean values with 68$\%$ confidence levels (CL), on both the free and select derived parameters of the $w_{sin}$CDM and $w_0w_a$CDM models for various datasets combinations.}
	\label{tab2}
	\scalebox{0.85}{
		\begin{tabular}{lccc}
			\hline
			\toprule
			\textbf{Dataset }&\;\;\;\;\;\textbf{DESI DR2+CC+Union3}\;\;\;\;\;& \textbf{DESI DR2+CC+PP} \;\;\;\;\;\;& \;\;\;\textbf{DESI DR2+CC+DES-5yr} 
			\\ \hline
			\textbf{Model} & \textbf{$\bm{w_{sin}}$CDM}\,&\textbf{$\bm{w_{sin}}$CDM}\,&\textbf{$\bm{w_{sin}}$CDM}\vspace{0.1cm}\\
			&\textcolor{teal}{\textbf{$w_0w_a$CDM}}\, & \textcolor{teal}{\textbf{$w_0w_a$CDM}}\, & \textcolor{teal}{\textbf{$w_0w_a$CDM}} 
			\\ \hline

			{\boldmath$10^{2}\omega_{b}$}&{$2.269\pm 0.037$ }&$2.271\pm 0.037$&$2.267\pm 0.037$\\
			
			&\textcolor{teal}{$2.266\pm 0.037 $} &\textcolor{teal}{$2.269\pm 0.037 $} &\textcolor{teal}{$2.264\pm 0.036$}\\

			{\boldmath$\omega{}_{\rm cdm }$}&$0.1234^{+0.0089}_{-0.0078} $ &$0.1194^{+0.0094}_{-0.0085}$ &$0.1247\pm 0.0080$\\
			
			&\textcolor{teal}{$0.1205\pm 0.0069$} &\textcolor{teal}{$0.1187\pm 0.0070 $} & \textcolor{teal}{$0.1220\pm 0.0067$}\\

			{\boldmath$H_0$ [$\text{km} \text{s}^{-1} \text{Mpc}^{-1}$]}&$66.8\pm 1.2$&$ 67.5\pm 1.3 $& $67.1\pm 1.1$\\
			
			& \textcolor{teal}{$66.7\pm 1.1 $}&\textcolor{teal}{$67.7\pm 1.2  $}  & \textcolor{teal}{$66.87\pm 0.95$}\\

			{\boldmath$w_{0}$} &$-0.72\pm 0.10$ &$-0.842^{+0.096}_{-0.11}$ &$-0.715\pm 0.068  $ \\
			&\textcolor{teal}{$-0.733\pm 0.094$}& \textcolor{teal}{$-0.877\pm 0.060$}& \textcolor{teal}{$-0.736\pm 0.065$} \\
			
			{\boldmath$w_{a}$} &$-0.89^{+0.49}_{-0.43}$ &$-0.42^{+0.48}_{-0.40}$ &$-0.95^{+0.40}_{-0.35} $ \\
			
			&\textcolor{teal}{$-0.78^{+0.45}_{-0.37}$}& \textcolor{teal}{$-0.33^{+0.36}_{-0.30}$}& \textcolor{teal}{$-0.86^{+0.37}_{-0.32}$}\\

			{\boldmath$M_B{\rm[mag]}$}&$-19.649\pm 0.096 $ &$-19.417^{+0.041}_{-0.036} $&$-19.537^{+0.036}_{-0.031}$ \\
			
			&$\textcolor{teal}{-19.656\pm 0.095}$ &\textcolor{teal}{$-19.420\pm 0.032$}& \textcolor{teal}{$-19.548\pm 0.029$}\\

			{\boldmath$\Omega{}_{m }  $}&$0.327\pm 0.014   $ &$0.312\pm 0.014 $&$0.327\pm 0.011  $\\
			
			&\textcolor{teal}{$0.323\pm 0.012$}& \textcolor{teal}{$0.312\pm 0.012 $} & \textcolor{teal}{$0.3240\pm 0.0096$}  \\

			{\boldmath$t{}_{0}  $}&$13.59^{+0.21}_{-0.26}$ &$ 13.69^{+0.23}_{-0.27}$  &$ 13.55^{+0.21}_{-0.24} $\\
			
			&\textcolor{teal}{$13.67\pm 0.19$}& \textcolor{teal}{$13.71\pm 0.19$} & \textcolor{teal}{$13.64\pm 0.19$}  \\

			\hline

			{\boldmath$ \chi^{2}_{min}$}&$ -22.19$&$-716.98$&$-838.26$ \\
			
			& \textcolor{teal}{$-23.26$}& \textcolor{teal}{$-718.15$} &\textcolor{teal}{$-839.41$}\\

			{\boldmath$\Delta \text{AIC}$}&$2.14$&$2.34$&$2.30$\\

			\hline
			\hline
		\end{tabular}
	}
\end{table*}

When the PP data are replaced by the DES-5yr data in the combined analysis (DR2+CC+DES-5yr), we observe from the third column of Table \ref{tab2} that the $H_0$ constraints obtained in our model are more consistent with Planck calibration measurements compared to those of the CPL model. The corresponding numerical values are summarized in the third row and third column of Table \ref{tab2}. Although the CPL model exhibits a slightly reduced $1\sigma$ error bar compared with our model, its central value remains less consistent with the Planck calibration. We also observe a strong deviation of $w_0$ from the cosmological constant at the $4.2\sigma$ level and a significant deviation of $w_a$ from zero at the $2.5\sigma$ level for both models. Moreover, the constraints on $\Omega_m$ are nearly identical to those obtained from the first data combination.\\

Further, we employed the Akaike Information Criterion (AIC) \cite{ref72} to compare our model with the CPL parametrization using different supernova samples, as summarized in Table \ref{tab2}. To quantify the relative performance of the two models, we adopted the $w_0w_a$CDM model as the reference scenario and set its AIC difference to zero. The resulting values of $\Delta \mathrm{AIC}$ values are presented in the last row of Table \ref{tab2}, where $\Delta \mathrm{AIC} = \mathrm{AIC}_{w_{\text{sin}}\mathrm{CDM}} - \mathrm{AIC}_{w_0w_a\mathrm{CDM}}$. We find that all $\Delta \mathrm{AIC}$ values are close to 2.3, indicating that our model is in good agreement with the observational data sets for all the data combinations considered in Table~\ref{tab2}.

\color{black}

\section{Conclusion}

In this study, we conducted a comprehensive observational analysis of the 
oscillatory dark energy model $w_{\sin}\mathrm{CDM}$, characterized by the 
equation of state
$$w_{\rm de}(a) = w_0 + w_a \left[ a \sin\!\left(\frac{1}{a}\right) - \sin(1) \right]$$
which naturally allows for periodic deviations from $w=-1$ at both low and 
intermediate redshifts. The oscillatory or periodic behavior observed in the $\omega_{sin}$CDM model can be theoretically motivated within multi-scalar field frameworks. In such scenarios, interactions among multiple scalar fields or suitable potentials can naturally lead to periodic features in the dark energy dynamics. From this perspective, the present phenomenological model may be viewed as an effective description of underlying multi-field theories \cite{ref77}.

Using the Friedmann framework for a flat FLRW Universe, we derived the corresponding dark-energy density evolution and constructed the Hubble expansion history $H(z)$ in terms of redshift. A joint likelihood analysis based on the latest DESI BAO (DR1), 33 OHD points, PantheonPlus, and SH0ES datasets was performed to obtain stringent constraints on both background and dynamical parameters.
A key result of our study is that the oscillatory parameters $(w_0,w_a)$ favour a mildly dynamical, quintessence-like dark energy behaviour across all dataset combinations, exhibiting a statistically significant deviation from the cosmological constant. In particular, PantheonPlus and SH0ES augmented combinations yielded
$$
w_0 = -0.85 \pm 0.08, \qquad 
w_a = -1.09^{+0.47}_{-0.42}$$
indicating that the dark energy may have undergone a phantom-crossing-like evolution in the past. Such features cannot be captured by the standard CPL model $w(a) = w_0 + w_a (1-a)$ or by the BA parametrization, both of which produce strictly monotonic EoS evolution. In contrast, the $w_{\sin}\mathrm{CDM}$ model naturally accommodates non-monotonic oscillatory behaviour, enhancing its flexibility in fitting multiple cosmological probes that peak at different redshifts. Compared with CPL or BA parametrizations, the oscillatory model shows significantly improved degeneracy breaking between $w_0$ and $w_a$ when DESI+PantheonPlus data are included, as reflected in the sharper confidence contours in the $w_0$-$w_a$ plane.

Our analysis further reveals that DESI's high-precision BAO measurements 
substantially strengthen the constraints on $\Omega_m$ and $H_0$, efficiently 
reducing parameter degeneracies that persist in CPL and related parametrizations. For the joint DESI+OHD+PantheonPlus+SH0ES dataset, the inferred Hubble constant is
$$
H_0 = 71.85 \pm 0.79 ~\mathrm{km\, s^{-1}\, Mpc^{-1}},$$
which reduces the Hubble tension to the $0.9\sigma$ level. This performance 
exceeds that of $\Lambda$CDM, for which the same dataset yields 
$H_0 = 70.97 \pm 0.56 ~\mathrm{km\, s^{-1}\, Mpc^{-1}}$, maintaining a 
tension of approximately $1.8\sigma$. The improvement in the oscillatory model originates from the additional freedom in the EoS, allowing adjustments in the expansion rate at intermediate redshifts ($0.5 \lesssim z \lesssim 2$), the regime where DESI carries the highest statistical weight. Simultaneously, the model preserves excellent agreement with the OHD and SNe~Ia evolution, as reflected in the reconstructed $H(z)$ and distance-modulus curves and matter density and age of the universe derived within the oscillatory model remain consistent with Planck-calibrated expectations and with the corresponding $\Lambda$CDM values. For the full dataset, we obtain
$$
\Omega_m = 0.321 \pm 0.011, \qquad 
t_0 = 13.01 \pm 0.17~\mathrm{Gyr},$$
demonstrating that introducing oscillations into the EoS does not disrupt the 
overall coherence of the late-time expansion history.\\

Furthermore, we carried out a detailed and comprehensive comparative analysis of our proposed
$w_{\rm sin}$CDM model and the CPL ($w_0w_a$CDM) parametrization by employing three different
combined data sets, namely DR2+OHD+Union3, DR2+OHD+PP, and DR2+OHD+DES-5yr. For all these data
combinations, the inferred values of the Hubble constant are found to lie in a relatively
narrow range, $H_0 \simeq 66.5$--$68~\mathrm{km\,s^{-1}\,Mpc^{-1}}$, indicating stable and
consistent constraints across different supernova samples. In addition, the present-day dark energy equation-of-state (EoS) parameter $w_0$ deviates from the cosmological constant value $w=-1$ at approximately the $2$--$4\sigma$
confidence level in both the $w_{\rm sin}$CDM and CPL models. This behavior suggests a mild
preference for dynamical dark energy over a strictly constant equation of state, although
the statistical significance depends on the chosen data combination. Our analysis also
consistently yields a comparatively higher present-day matter density parameter,
$\Omega_m \approx 0.32$, for both models across all considered data sets. The close similarity of the inferred cosmological parameters in the two frameworks indicates that current background observations do not strongly discriminate between the $w_{\rm sin}$CDM and CPL models, although both provide a comparably good description of the observational data.\\

Overall, our results indicate that the $w_{\sin}\mathrm{CDM}$ parametrization 
provides a viable and competitive alternative to the standard CPL/BA parametrizations and to $\Lambda$CDM. The oscillatory behavior of the EoS not only accommodates physically motivated dynamical deviations from $w=-1$ but also offers a moderate alleviation of the Hubble tension without introducing inconsistencies with other cosmological probes. With future high-precision BAO measurements from DESI, along with improved standard-siren and cosmic-chronometer data, oscillatory dark-energy models such as $w_{\sin}\mathrm{CDM}$ will become increasingly testable. This study underscores the broader importance of non-monotonic, physically motivated EoS models as promising frameworks for understanding late-time cosmic acceleration and addressing persistent cosmological tensions.

\section*{Declaration of competing interest}
	\noindent 
We wish to confirm that there are no known conflicts of interest
associated with this publication and that there has been no significant financial
support for this work that could have influenced its outcome.

\section*{Data availability}
	\noindent 
We employed the publicly available DESI BAO data, Pantheon Plus (PP) data and Observational Hubble Parameter (OHD) data presented in this study. DESI BAO data are accessible from the official repository at https://data.desi.lbl.gov/doc/releases/. The OHD data were compiled from publicly available cosmic chronometer measurements in the literature, with a representative compilation available at: https://github.com/AhmadMehrabi Cosmic chronometer data. The Pantheon Plus (PP) compilation (distance moduli and covariance matrices), is publicly available on the GitHub:https://github.com/brinckmann/montepythonpublic/tree/3.6/montepython/likelihoods/Pantheon Plus. No additional data were used in this study.

\begin{center}
   \textbf{Appendix I : Triangle Countor  } 
\end{center}

In this appendix, we present triangular plots with 1D- and 2D- marginalized distributions for all considered parameters presented in Table \ref{tab1}  within the $w_{\rm{sin}}$CDM and $\Lambda$CDM models based on all given combinations (See Figures 8 and 9). 
\begin{figure}[hbt!]
    \centering
    \includegraphics[width=0.8\linewidth]{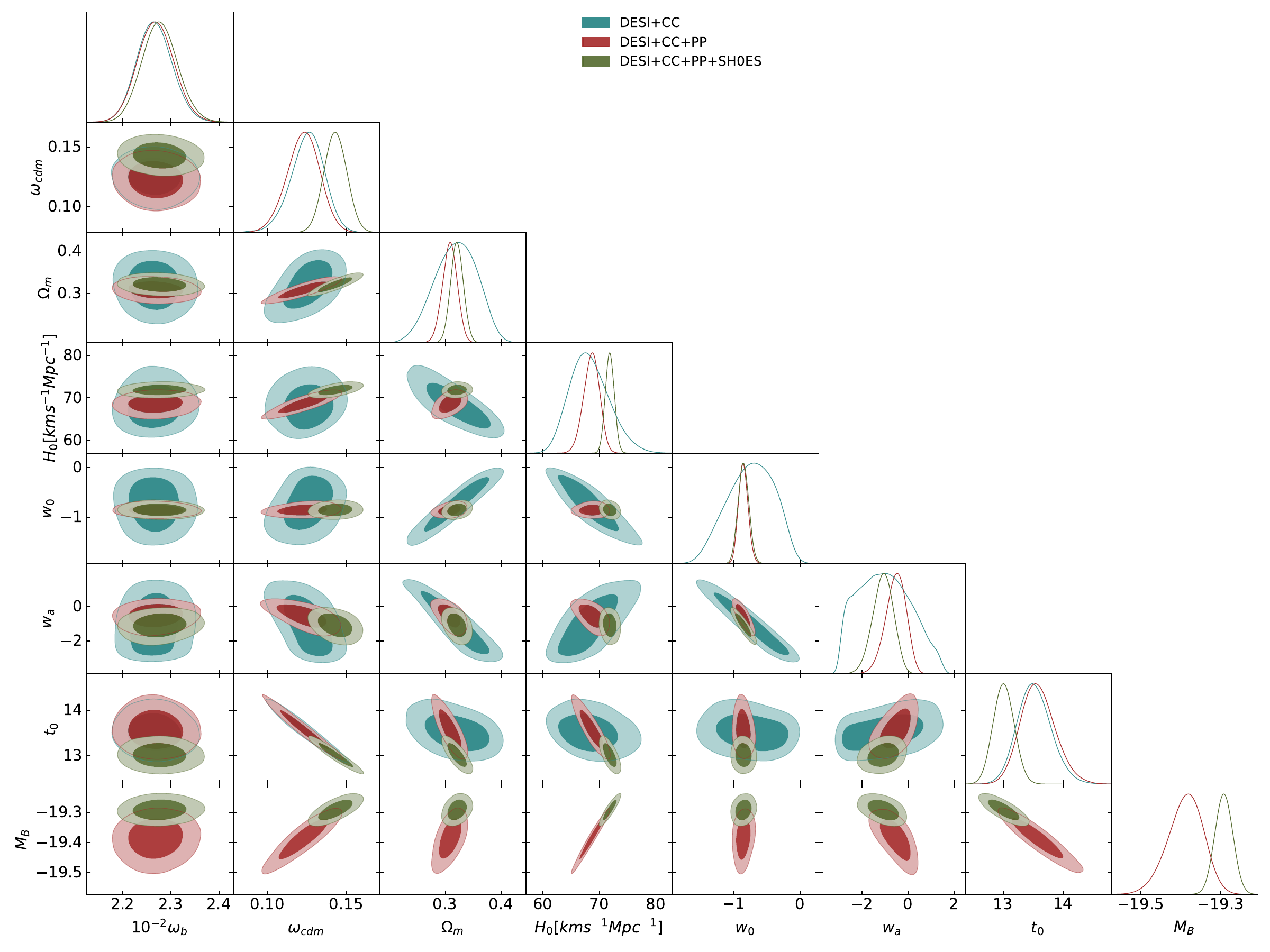}
    \caption{In this triangular plot, we present 1D- and 2D- marginalized distributions for all considered parameters presented in Table \ref{tab1}  within the $w_{\rm{sin}}$CDM model based on all consider data sets combinations. }
    \label{fig8}
\end{figure}
\pagebreak

\begin{figure}[hbt!]
    \centering
    \includegraphics[width=0.8\linewidth]{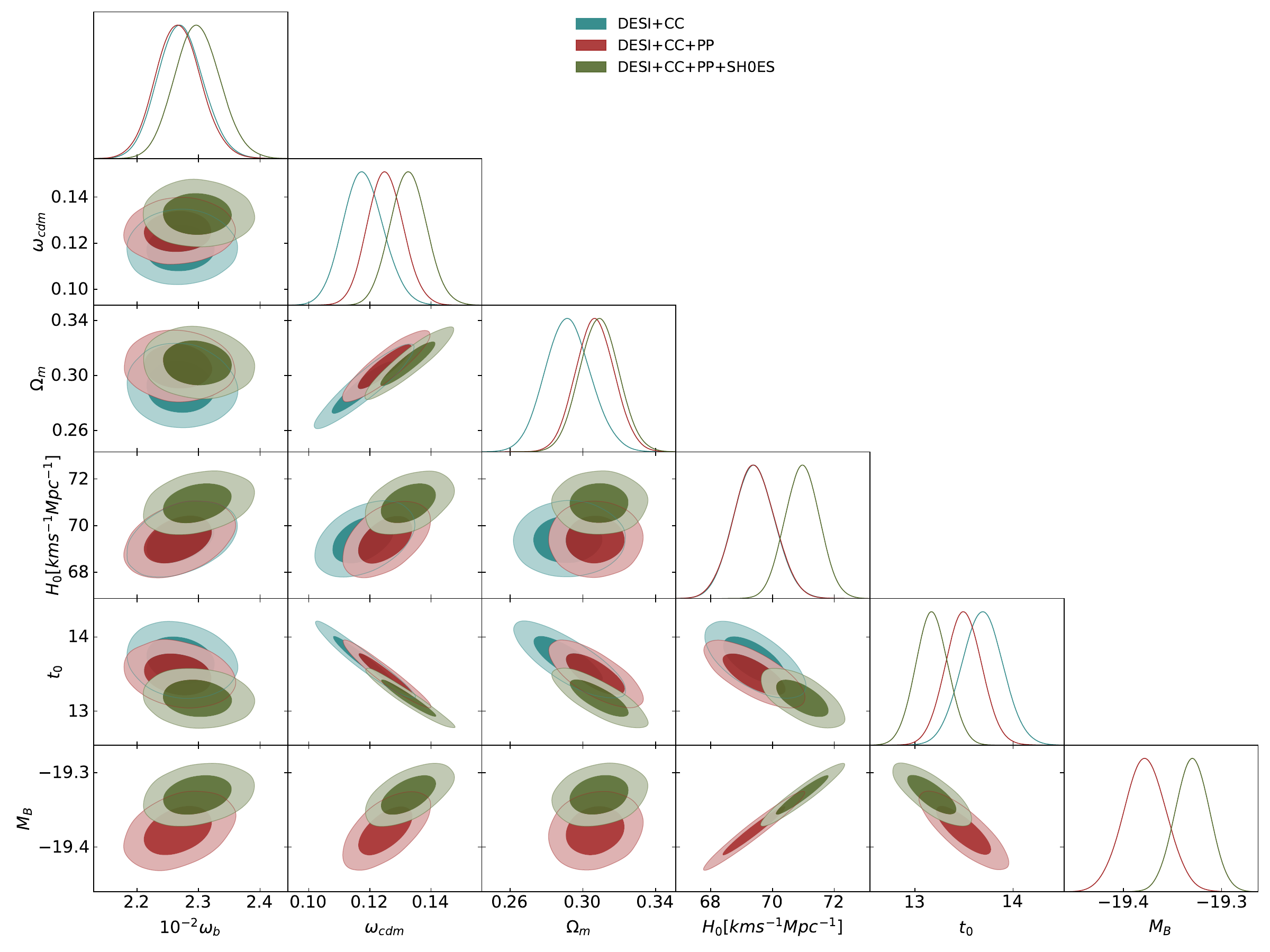}
    \caption{ In this triangular plot, we present 1D- and 2D- marginalized distributions for all considered parameters presented in Table \ref{tab1}  within the $\Lambda$CDM model based on all consider datasets combinations}
    \label{fig9}
\end{figure}

\begin{acknowledgments}
\noindent  
 The authors (A. Dixit and A. Pradhan) are thankful to IUCAA, Pune, India for providing support and facilities under the Visiting Associateship program. M. Yadav was sponsored by a senior Research Fellowship from the Council of Scientific and Industrial Research, Government of India (CSIR/UGC Ref.\ No.\ 180010603050). The authors appreciate  the Reviewer and Editor for their informative remarks, which improved the manuscript in its current form. 
\end{acknowledgments}

\end{document}